\documentclass[aps,pra,preprint,onecolumn,notitlepage,showkeys,nofootinbib]{revtex4-2}
\usepackage[T1]{fontenc}
\usepackage{amsmath,amssymb,graphicx}
\usepackage{amsthm}
\usepackage{booktabs}
\usepackage{bm}
\usepackage{hyperref}

\makeatletter
\def\frontmatter@abstract@produce{%
  \par
  \addvspace{\frontmatter@preabstractspace}%
  \begingroup
   \dimen@\baselineskip
   \setbox\z@\vtop{\unvcopy\absbox}%
   \advance\dimen@-\ht\z@\advance\dimen@-\prevdepth
   \@ifdim{\dimen@>\z@}{\vskip\dimen@}{}%
  \endgroup
  \begingroup
   \prep@absbox
   \unvbox\absbox
   \post@absbox
  \endgroup
 \@ifx{\@empty\mini@notes}{}{\mini@notes\par}%
 \addvspace\frontmatter@postabstractspace
}%
\makeatother

\newtheoremstyle{normalstyle}{6pt}{6pt}{\normalfont}{}{\normalfont}{.}{0.5em}{}
\theoremstyle{normalstyle}
\newtheorem{lemma}{Lemma}[section]
\newtheorem{theorem}[lemma]{Theorem}
\newtheorem{corollary}[lemma]{Corollary}
\newenvironment{proofplain}{\par\noindent Proof.\ }{\hfill$\square$\par}

\newcommand{\id}{\mathbb{I}}
\newcommand{\HH}{\mathcal{H}}
\newcommand{\LL}{\mathcal{L}}

\newcommand{\fstar}{f_{\star}}
\newcommand{\THaw}{T_{\mathrm H}}
\newcommand{\Ftel}{F_{\mathrm{tel}}}

\newcommand{\paperabstract}{%
Fermionic teleportation near a Schwarzschild horizon involves two physically distinct ingredients that are often varied together: Hawking thermal mixing and the relative composition of the right and left Unruh modes. We separate them by factorizing the reduced map of the detected exterior fermionic mode into a mode-preparation channel and a Hawking thermal channel. At fixed right-left Unruh-mode composition, we then solve a global constrained quantum-channel optimization over all two-qubit resource states whose fully entangled fraction remains at or below the standard teleportation threshold immediately before thermal mixing. A parity twirl and positivity reduce the arbitrary-state problem exactly to a two-variable quadratic optimization. The solution yields a necessary-and-sufficient activation boundary, a second exact boundary for attaining the universal optimum, and explicit optimizing states. The resulting phase diagram contains a strict no-go region, a composition-limited activation region, and a universal-bound region. The analysis converts state-specific observations of Hawking-assisted teleportation into exact channel-level bounds over the full two-qubit state space.}

\begin{document}

\title{Exact teleportation bounds for fermionic Hawking thermality beyond the single-mode approximation}
\author{Zhiming Huang}
\email{465609785@qq.com}
\affiliation{School of Economics and Management, Wuyi University, Jiangmen 529020, China}

\begin{abstract}
\paperabstract
\end{abstract}

\maketitle

\section{Introduction}

Quantum teleportation turns shared entanglement into an operational communication resource \cite{Bennett1993}.  For two qubits, the fully entangled fraction (FEF)
\begin{equation}
 f(\rho)=\max_{\Phi}\langle\Phi|\rho|\Phi\rangle,
 \label{eq:fef-def}
\end{equation}
where the maximization is over maximally entangled states, determines the standard average teleportation fidelity through
\begin{equation}
 \Ftel=\frac{2f+1}{3}.
 \label{eq:tele-fidelity}
\end{equation}
Thus $f>1/2$ is the familiar threshold for beating the classical average-fidelity benchmark $2/3$ in the standard two-qubit setting \cite{Horodecki1999}.  Local dissipative dynamics can sometimes increase the FEF of suitable mixed states and can even move a state across this threshold \cite{Badziag2000,Yeo2008}.  The relevant question for a specified physical channel is therefore quantitative: how large can the enhancement be, which states attain it, and in which parameter region is threshold crossing impossible?

Relativistic quantum information adds a further physical distinction.  Entanglement and teleportation have been studied for accelerated observers and black-hole spacetimes since the early analyses of inaccessible field modes \cite{FuentesSchuller2005,PanJing2008,MartinMartinez2010}.  Beyond the single-mode approximation, the coefficients of the right and left Unruh modes encode the chosen mode or wave-packet preparation rather than an additional thermal parameter \cite{Bruschi2010}.  More generally, optimization over noisy qubit channels shows that non-unital channels may favor nonmaximally entangled inputs \cite{Pal2014}, and the Unruh effect itself admits a quantum-channel description \cite{Banerjee2017}.  For Dirac fields, special initial states can display increasing or nonmonotonic teleportation fidelity under acceleration or horizon-induced mode mixing \cite{ChenChan2019}.  In Schwarzschild spacetime, Wu et al. found decreasing, increasing, and nonmonotonic FEF curves with Hawking temperature and also examined the dependence on the relative composition of the right and left Unruh modes, parametrized by $q_R$ \cite{Wu2023}.  Related extensions include three-qubit FEF and black-hole models with decoherence or local filtering \cite{Mi2025,LiuLongHe2025}; recent work on two-qubit faithfulness likewise distinguishes the behavior of unital and non-unital channels \cite{LiuLi2025}.

The unresolved issue is that two effects are often varied together.  The Hawking temperature controls thermal mixing, whereas the amplitudes $q_R$ and $q_L$, constrained by $|q_R|^2+|q_L|^2=1$, control the relative composition of the right and left Unruh modes.  Varying both simultaneously can therefore obscure which part of an observed change is genuinely thermal.  In addition, existing black-hole teleportation studies are mostly state specific: selected input states are evolved and their FEF curves are inspected, but this does not determine the global optimum over all admissible two-qubit resources or the exact region where thermal activation is impossible.

In this work we formulate the problem as an exact quantum-channel optimization.  Starting from the fermionic Schwarzschild mode transformation, we factorize the detected exterior-mode map into a mode-preparation channel and a Hawking thermal channel.  We then fix the relative composition of the right and left Unruh modes and maximize the output FEF over all two-qubit resources that remain at or below the standard threshold immediately before thermal mixing.  The optimization over arbitrary two-qubit density matrices is reduced exactly to a two-variable quadratic problem.  This yields a necessary-and-sufficient activation boundary, a universal thermal upper bound, a second boundary for attaining that bound, explicit optimizing states, and a closed operational phase diagram with a strict no-go region, a composition-limited activation region, and a universal-bound region.

The paper is organized as follows.  Section~\ref{sec:channel} derives the factorized horizon map and explains its physical meaning.  Section~\ref{sec:universal} establishes the universal Hawking thermal bound.  Section~\ref{sec:fixedq} solves the constrained global optimization at fixed relative composition of the right and left Unruh modes.  Section~\ref{sec:phase} gives the operational phase diagram and its limiting regimes.  Section~\ref{sec:comparison} compares the exact bounds with representative state-specific results and discusses their physical interpretation.  Section~\ref{sec:conclusion} summarizes the main results and conclusions.

\section{Fermionic Schwarzschild map and channel factorization}
\label{sec:channel}

Consider the Schwarzschild geometry
\begin{equation}
 ds^2=-\left(1-\frac{2M}{r}\right)dt^2
 +\left(1-\frac{2M}{r}\right)^{-1}dr^2
 +r^2d\Omega_2^2,
 \label{eq:schwarzschild}
\end{equation}
with Hawking temperature
\begin{equation}
 \THaw=\frac{1}{8\pi M}.
 \label{eq:TH}
\end{equation}
We use natural units.  For a massless Dirac mode of Killing frequency $\omega$, define
\begin{equation}
 \eta=\frac{1}{1+e^{-\omega/\THaw}},
 \qquad
 s=\sqrt{\eta}.
 \label{eq:eta-s}
\end{equation}
For positive Hawking temperature, $0<e^{-\omega/\THaw}<1$, so $1/2<\eta<1$ and hence $1/\sqrt2<s<1$; the endpoints are reached only in the formal high-temperature and zero-temperature limits.  The right and left Unruh amplitudes obey
\begin{equation}
 |q_R|^2+|q_L|^2=1.
 \label{eq:q-normalization}
\end{equation}
After the inaccessible sectors are traced out, the phase of $q_L$ drops out and the phase of $q_R$ can be absorbed into the basis of the detected exterior fermionic mode.  Consequently the FEF depends on the relative composition of the right and left Unruh modes only through the single real parameter
\begin{equation}
 q=|q_R|\in[0,1],
 \qquad |q_L|=\sqrt{1-q^2}.
 \label{eq:q-def}
\end{equation}

Tracing out the inaccessible interior modes and the exterior antifermion gives the following map on Bob's qubit:
\begin{equation}
 \Lambda_{\eta,q}
 \begin{pmatrix}
 x&z\\ z^*&y
 \end{pmatrix}
 =
 \begin{pmatrix}
 \eta[x+(1-q^2)y]&q\sqrt\eta\,z\\
 q\sqrt\eta\,z^*&(1-\eta)x+[1-(1-q^2)\eta]y
 \end{pmatrix}.
 \label{eq:full-map}
\end{equation}
Appendix~\ref{app:field} derives Eq.~\eqref{eq:full-map} directly from the fermionic Unruh vacuum and one-particle excitation.

The first structural result is an exact factorization.

\begin{theorem}[Horizon-map factorization]
The reduced fermionic Schwarzschild map in Eq.~\eqref{eq:full-map} can be written as
\begin{equation}
 \Lambda_{\eta,q}=\HH_{\eta}\circ\LL_q,
 \label{eq:factorization}
\end{equation}
where
\begin{equation}
 \LL_q
 \begin{pmatrix}
 x&z\\z^*&y
 \end{pmatrix}
 =
 \begin{pmatrix}
 x+(1-q^2)y&qz\\qz^*&q^2y
 \end{pmatrix}
 \label{eq:L-map}
\end{equation}
and
\begin{equation}
 \HH_{\eta}
 \begin{pmatrix}
 x&z\\z^*&y
 \end{pmatrix}
 =
 \begin{pmatrix}
 \eta x&\sqrt\eta\,z\\
 \sqrt\eta\,z^*&(1-\eta)x+y
 \end{pmatrix}.
 \label{eq:H-map}
\end{equation}
\end{theorem}

\begin{proofplain}
The Kraus operators can be constructed directly from the population transfer and coherence damping in Eqs.~\eqref{eq:L-map} and \eqref{eq:H-map}.  For $\LL_q$, the diagonal operator $L_0$ keeps the $|0\rangle$ amplitude unchanged while retaining only a factor $q$ of the $|1\rangle$ amplitude; the missing excited-state probability $1-q^2$ must therefore be transferred from $|1\rangle$ to $|0\rangle$ by a second operator.  This motivates the choice
\begin{equation}
 L_0=\begin{pmatrix}1&0\\0&q\end{pmatrix},\quad
 L_1=\begin{pmatrix}0&\sqrt{1-q^2}\\0&0\end{pmatrix}.
 \label{eq:L-kraus}
\end{equation}
They satisfy
\begin{equation}
 L_0^\dagger L_0+L_1^\dagger L_1
 =\begin{pmatrix}1&0\\0&q^2\end{pmatrix}
 +\begin{pmatrix}0&0\\0&1-q^2\end{pmatrix}
 =\id,
 \label{eq:L-completeness}
\end{equation}
so the map is trace preserving.  Moreover,
\begin{equation}
 \sum_{j=0}^{1}L_j
 \begin{pmatrix}x&z\\z^*&y\end{pmatrix}
 L_j^\dagger
 =
 \begin{pmatrix}x+(1-q^2)y&qz\\qz^*&q^2y\end{pmatrix},
\end{equation}
which is exactly Eq.~\eqref{eq:L-map}; the operator-sum representation also guarantees complete positivity.

For $\HH_\eta$, the $|0\rangle$ amplitude is retained with factor $\sqrt\eta$ while $|1\rangle$ is unchanged by the no-transition part.  The lost probability $1-\eta$ from $|0\rangle$ appears in the $|1\rangle$ population, so a second Kraus operator transfers $|0\rangle$ to $|1\rangle$.  We therefore choose
\begin{equation}
 H_0=\begin{pmatrix}\sqrt\eta&0\\0&1\end{pmatrix},\quad
 H_1=\begin{pmatrix}0&0\\\sqrt{1-\eta}&0\end{pmatrix}.
 \label{eq:H-kraus}
\end{equation}
Here
\begin{equation}
 H_0^\dagger H_0+H_1^\dagger H_1
 =\begin{pmatrix}\eta&0\\0&1\end{pmatrix}
 +\begin{pmatrix}1-\eta&0\\0&0\end{pmatrix}
 =\id,
 \label{eq:H-completeness}
\end{equation}
and
\begin{equation}
 \sum_{j=0}^{1}H_j
 \begin{pmatrix}x&z\\z^*&y\end{pmatrix}
 H_j^\dagger
 =
 \begin{pmatrix}\eta x&\sqrt\eta\,z\\
 \sqrt\eta\,z^*&(1-\eta)x+y\end{pmatrix},
\end{equation}
which reproduces Eq.~\eqref{eq:H-map}.  Thus both factors are completely positive and trace preserving.  Composing the two matrices gives Eq.~\eqref{eq:full-map}, proving Eq.~\eqref{eq:factorization}.
\end{proofplain}

The factorization fixes the physical interpretation.  The parameter $q=|q_R|$, together with $|q_L|^2=1-q^2$, specifies the relative composition of the right and left Unruh modes in the prepared excitation; beyond the single-mode approximation, this composition is tied to the chosen mode or wave packet \cite{Bruschi2010}.  In contrast, all dependence on $\THaw/\omega$ is carried by $\HH_\eta$.  Thus different $q$ values should not be interpreted as different strengths of Hawking noise.  They define different accessible input resources for the same thermal map.

In the Bloch representation, the full map takes
\begin{equation}
 (r_x,r_y,r_z)\mapsto
 \left(qs\,r_x,\;qs\,r_y,\;q^2s^2r_z+s^2(2-q^2)-1\right).
 \label{eq:bloch-full}
\end{equation}
The translation term vanishes along
\begin{equation}
 s^2(2-q^2)=1,
 \label{eq:unital-line}
\end{equation}
which is the unital line of the combined channel.  This observation is consistent with the general fact that unital local channels cannot enhance the relevant two-qubit fidelity-based faithfulness \cite{Badziag2000,LiuLi2025}.  The main optimization below, however, isolates the thermal map at fixed relative composition of the right and left Unruh modes rather than optimizing the combined map over $q$.

\section{Universal bound for Hawking thermality}
\label{sec:universal}

Let
\begin{equation}
 \sigma=(\id\otimes\LL_q)(\rho_{\rm prep})
 \label{eq:accessible-resource}
\end{equation}
be the resource actually accessible before thermal Hawking mixing.  We impose
\begin{equation}
 f(\sigma)\le\frac12.
 \label{eq:threshold-constraint}
\end{equation}
The operational question is whether $\HH_\eta$ can move such a resource above the standard FEF threshold, and if so by how much.

Before retaining the fixed-$q$ image constraint in Eq.~\eqref{eq:accessible-resource}, it is useful to solve the larger optimization over all two-qubit $\sigma$ satisfying Eq.~\eqref{eq:threshold-constraint}.

\begin{theorem}[Universal thermal bound]\label{thm:universal}
For every two-qubit state $\sigma$ with $f(\sigma)\le1/2$,
\begin{equation}
 f\!\left[(\id\otimes\HH_{\eta})(\sigma)\right]
 \le
 \frac12+\frac{s^2(1-s)}{2(1+3s)}.
 \label{eq:universal-bound}
\end{equation}
The bound is attainable for every $1/\sqrt2\le s<1$.
\end{theorem}

\begin{proofplain}
Choose the maximizing output maximally entangled state and use a local unitary on Alice to map it to $|\Phi^+\rangle=(|00\rangle+|11\rangle)/\sqrt2$.  Because the channel acts only on Bob, this does not change the optimization.  A $Z\otimes Z$ parity twirl preserves the $|\Phi^+\rangle$ output overlap and cannot increase the input FEF, so an optimizer exists among $X$ states.  The remaining reduction is given in Appendix~\ref{app:reduction}.  For the thermal map alone it leaves the one-dimensional boundary problem
\begin{equation}
 2f_{\rm out}(x)
 =1+2s(1-s)x-(1-s)(1+3s)x^2,
 \qquad 0\le x\le1,
 \label{eq:univ-parabola}
\end{equation}
For $s<1$, the coefficient of $x^2$ is $-(1-s)(1+3s)<0$, so this is a strictly concave parabola.  Differentiating gives
\begin{equation}
 \frac{d(2f_{\rm out})}{dx}
 =2(1-s)\left[s-(1+3s)x\right],
\end{equation}
and hence the unique stationary point is
\begin{equation}
 x_*=\frac{s}{1+3s}.
 \label{eq:xstar}
\end{equation}
Because $0<x_*<1$, it lies inside the allowed interval and is the global maximum.  Substitution into Eq.~\eqref{eq:univ-parabola} gives Eq.~\eqref{eq:universal-bound}.
\end{proofplain}

An explicit state that attains Eq.~\eqref{eq:universal-bound} is
\begin{equation}
 \sigma_*(s)=
 \begin{pmatrix}
 x_*^2&0&0&x_*(1-x_*)\\
 0&0&0&0\\
 0&0&2x_*(1-x_*)&0\\
 x_*(1-x_*)&0&0&(1-x_*)^2
 \end{pmatrix},
 \label{eq:sigma-star}
\end{equation}
Its trace is one.  The nontrivial $|00\rangle$--$|11\rangle$ block has zero determinant and the remaining nonzero diagonal entry is $2x_*(1-x_*)\ge0$, so $\sigma_*$ is a valid density matrix.  Direct evaluation gives $f(\sigma_*)=1/2$.  The optimizer is therefore a threshold resource, not a maximally entangled state.

In the formal high-temperature fermionic limit $s\to1/\sqrt2$,
\begin{equation}
 f_{\rm H,max}
 =\frac{9+4\sqrt2}{28}
 \approx0.5234590803,
 \label{eq:highT-fmax}
\end{equation}
and Eq.~\eqref{eq:tele-fidelity} gives
\begin{equation}
 \Ftel^{\rm H,max}
 =\frac{23+4\sqrt2}{42}
 \approx0.6823060536.
 \label{eq:highT-Ftel}
\end{equation}
At low temperature, with $\epsilon=e^{-\omega/\THaw}$,
\begin{equation}
 f_{\rm H,max}-\frac12
 =\frac{\epsilon}{16}+O(\epsilon^2),
 \qquad
 \Ftel^{\rm H,max}-\frac23
 =\frac{\epsilon}{24}+O(\epsilon^2).
 \label{eq:lowT-expansion}
\end{equation}
Thus the existence of an activated resource at any nonzero temperature does not imply a large effect for $\THaw\ll\omega$.

\section{Fixed relative composition of the right and left Unruh modes: exact constrained optimum}
\label{sec:fixedq}

The universal state in Eq.~\eqref{eq:sigma-star} need not belong to the image of a fixed $\LL_q$.  The physically sharper problem is therefore
\begin{equation}
 \fstar(q,s)=
 \max_{\substack{\rho_{\rm prep}\ge0,\;\mathrm{Tr}\rho_{\rm prep}=1\\
 f[(\id\otimes\LL_q)(\rho_{\rm prep})]\le1/2}}
 f\!\left[(\id\otimes\HH_\eta\LL_q)(\rho_{\rm prep})\right].
 \label{eq:fixedq-problem}
\end{equation}
Appendix~\ref{app:reduction} proves that this optimization over arbitrary two-qubit density matrices is exactly equivalent to maximizing
\begin{equation}
 2f_{\rm out}(x,y)
 =1-s^2+(2s^2-1)x^2+s^2q^2y^2+2q sxy
 \label{eq:xy-objective}
\end{equation}
over
\begin{equation}
 x\ge0,\qquad y\ge0,
 \qquad x^2+y^2\le1,
 \qquad x+qy\le1.
 \label{eq:xy-domain}
\end{equation}
The disk is the positivity and normalization boundary of the optimized preparation state, while the line is precisely the constraint that the accessible resource remains at or below $f=1/2$.

Let
\begin{equation}
 Q=q^2,
 \qquad
 s_c=\frac{1+\sqrt{33}}{8}.
 \label{eq:Q-sc}
\end{equation}
The first branch boundary is
\begin{equation}
 Q_1(s)=
 \frac{3s^2-3s-1+
 \sqrt{(1-s)(-s^3+11s^2+5s+1)}}{s(2s-1)},
 \label{eq:Q1}
\end{equation}
when $1/\sqrt2\le s<s_c$, and $Q_1(s)=0$ for $s\ge s_c$.  The second branch boundary is
\begin{equation}
 Q_2(s)=\frac{1+2s}{1+4s}.
 \label{eq:Q2}
\end{equation}
The algebra leading to these boundaries is collected in Appendix~\ref{app:branches}.

\begin{theorem}[Exact optimum at fixed relative composition of the right and left Unruh modes]
For $1/\sqrt2\le s<1$ and $0\le q\le1$, the solution of Eq.~\eqref{eq:fixedq-problem} is
\begin{equation}
 \fstar(q,s)=
 \begin{cases}
 f_{\rm I}(q,s),&0\le Q\le Q_1(s),\\[1mm]
 f_{\rm II}(q,s),&Q_1(s)<Q<Q_2(s),\\[1mm]
 f_{\rm III}(s),&Q_2(s)\le Q\le1,
 \end{cases}
 \label{eq:piecewise}
\end{equation}
where
\begin{equation}
 f_{\rm I}
 =\frac{1+s^2Q+
 \sqrt{[s^2(2-Q)-1]^2+4s^2Q}}{4},
 \label{eq:fI}
\end{equation}
\begin{equation}
 f_{\rm II}
 =\frac12+
 \frac{(1-s)(1-Q)[Q(1+5s)-(1+s)]}{2(1+Q)^2},
 \label{eq:fII}
\end{equation}
and
\begin{equation}
 f_{\rm III}
 =\frac12+\frac{s^2(1-s)}{2(1+3s)}.
 \label{eq:fIII}
\end{equation}
At $s=1$ the thermal map is the identity and the constrained optimum is $1/2$.
\end{theorem}

\begin{proofplain}
In the first quadrant, all coefficients multiplying $x^2$, $y^2$, and $xy$ in Eq.~\eqref{eq:xy-objective} are nonnegative.  Hence increasing either variable cannot decrease the objective, so a maximum must lie on the upper boundary of the feasible set: either the circular boundary $x^2+y^2=1$ or the threshold line $x+qy=1$.  On the circle, the quadratic part is determined by a symmetric $2\times2$ matrix with nonnegative off-diagonal entry; its largest eigenvector can be chosen in the first quadrant, and its largest eigenvalue gives Eq.~\eqref{eq:fI}.  This is branch I as long as that eigenvector still satisfies $x+qy\le1$.

When the circular maximizer first reaches the threshold line, both constraints are active.  Solving $x^2+y^2=1$ together with $x+qy=1$ gives
\begin{equation}
 x_c=\frac{1-Q}{1+Q},
 \qquad
 y_c=\frac{2q}{1+Q},
 \label{eq:circle-line-point}
\end{equation}
and substitution yields Eq.~\eqref{eq:fII}.  The value of $Q$ at which the circular maximizer becomes tangent to the line is precisely $Q_1(s)$.

On the threshold line, substituting $y=(1-x)/q$ makes the coefficient of $x^2$ equal to $3s^2-2s-1=(3s+1)(s-1)<0$ for $s<1$, so the line-restricted objective is concave.  Its stationary point is
\begin{equation}
 x_*=\frac{s}{1+3s},
 \qquad
 qy_*=\frac{1+2s}{1+3s}.
 \label{eq:line-stationary}
\end{equation}
If this point lies outside the disk, the best feasible point on the line is its endpoint at the circle-line intersection, which is branch II.  It enters the disk exactly when $Q\ge Q_2(s)$; from then on it gives branch III and Eq.~\eqref{eq:fIII}.  Continuity at $Q_1$ and $Q_2$ follows by direct substitution.
\end{proofplain}

The middle and last branches admit simple optimizing preparations.  In the middle branch the preparation can be chosen pure,
\begin{equation}
 |\psi_{\rm II}\rangle
 =\frac{1-Q}{1+Q}|00\rangle
 +\frac{2q}{1+Q}|11\rangle.
 \label{eq:psi-II}
\end{equation}
The two amplitudes are normalized because their squared moduli sum to one; this is exactly the circle-line intersection in Eq.~\eqref{eq:circle-line-point}.  For $Q\ge Q_2(s)$, an optimizer is
\begin{equation}
 \rho_{\rm III}=
 \begin{pmatrix}
 x_*^2&0&0&x_*y_*\\
 0&0&0&0\\
 0&0&1-x_*^2-y_*^2&0\\
 x_*y_*&0&0&y_*^2
 \end{pmatrix},
 \label{eq:rho-III}
\end{equation}
with $x_*$ and $y_*$ from Eq.~\eqref{eq:line-stationary}.  Its $|00\rangle$--$|11\rangle$ block is rank one, while the remaining diagonal entry is nonnegative precisely because $x_*^2+y_*^2\le1$.  Hence it is a valid density matrix.  In both branches the accessible state after $\LL_q$ lies exactly on the threshold $f=1/2$.

\section{Operational phase diagram}
\label{sec:phase}

Although the full optimum contains the branch $Q_1$, activation across the standard threshold is controlled by a simpler boundary.  Define
\begin{equation}
 Q_0(s)=\frac{1+s}{1+5s}.
 \label{eq:Q0}
\end{equation}
For the physical interval $1/\sqrt2\le s<1$, one has
\begin{equation}
 Q_1(s)<Q_0(s)<Q_2(s)
 \label{eq:ordering}
\end{equation}
whenever $Q_1$ is nonzero; Appendix~\ref{app:branches} gives a direct proof.

\begin{corollary}[Necessary and sufficient thermal-activation condition]
For fixed $q$ and finite nonzero Hawking temperature, there exists a preparation state whose accessible resource satisfies $f\le1/2$ before thermal mixing and $f>1/2$ after thermal mixing if and only if
\begin{equation}
 q^2>\frac{1+s}{1+5s}.
 \label{eq:activation-condition}
\end{equation}
\end{corollary}

\begin{proofplain}
Branch I lies entirely below the activation boundary because $Q_1<Q_0$; direct substitution of its admissible range into Eq.~\eqref{eq:fI} gives $f_{\rm I}\le1/2$.  On branch II, Eq.~\eqref{eq:fII} factorizes as
\begin{equation}
 f_{\rm II}-\frac12
 =\frac{(1-s)(1-Q)[Q(1+5s)-(1+s)]}{2(1+Q)^2}.
 \label{eq:fII-factor}
\end{equation}
For $s<1$ and $Q<1$, every factor outside the final square bracket is nonnegative, so the sign is exactly the sign of $Q(1+5s)-(1+s)$.  Hence branch II crosses the threshold precisely at $Q=Q_0(s)$.  On branch III, $f_{\rm III}-1/2=s^2(1-s)/[2(1+3s)]>0$ for $s<1$, and Eq.~\eqref{eq:ordering} gives $Q_2>Q_0$.  Therefore strict activation occurs if and only if $Q>Q_0(s)$, which is Eq.~\eqref{eq:activation-condition}.
\end{proofplain}

\begin{figure}[t]
\centering
\includegraphics[width=0.82\textwidth]{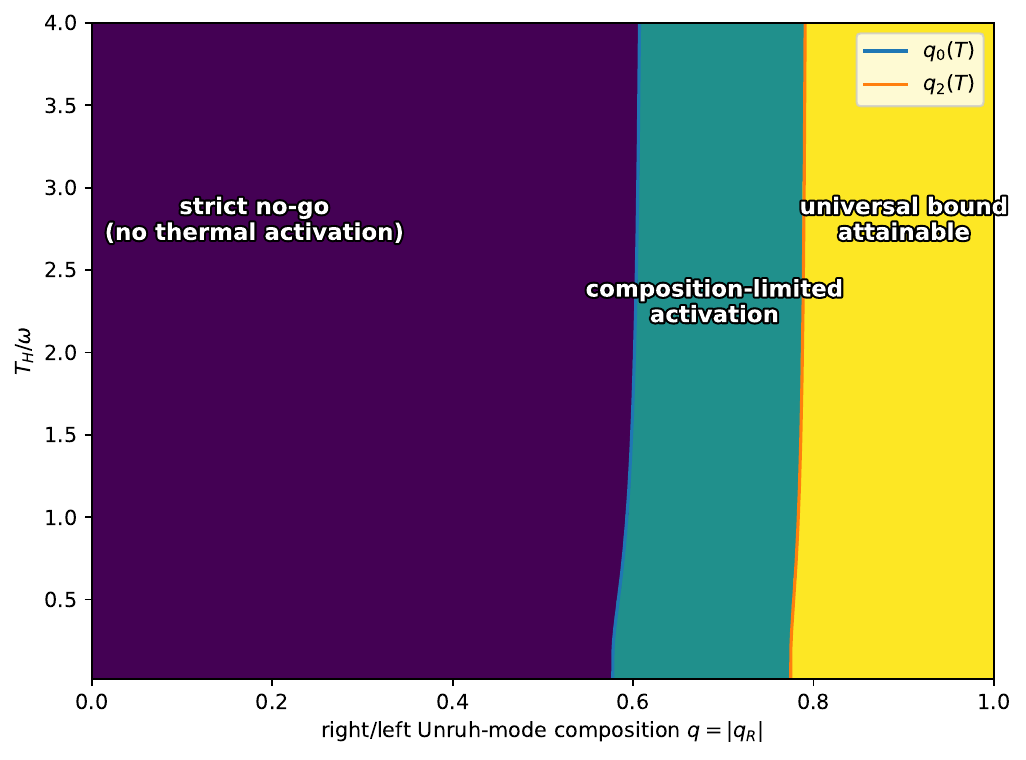}
\caption{Exact operational phase diagram at fixed relative composition of the right and left Unruh modes.  The lower boundary $q_0=\sqrt{Q_0}$ is necessary and sufficient for a strict crossing of the standard FEF threshold under the thermal Hawking map.  Above $q_2=\sqrt{Q_2}$ the universal bound in Eq.~\eqref{eq:universal-bound} is attainable.  The plotted parameter $q=|q_R|$ parametrizes the relative composition of the right and left Unruh modes through $|q_R|^2+|q_L|^2=1$; it is not a second Hawking temperature.}
\label{fig:phase}
\end{figure}

A second operational boundary is Eq.~\eqref{eq:Q2}: above it a fixed relative composition of the right and left Unruh modes can realize the universal state of Eq.~\eqref{eq:sigma-star}, while below it the image of $\LL_q$ limits the attainable enhancement.  This gives three regions:
\begin{equation}
 \begin{array}{ll}
 Q\le Q_0(s):&\text{strict no-go (no thermal activation)},\\[1mm]
 Q_0(s)<Q<Q_2(s):&\text{composition-limited thermal activation},\\[1mm]
 Q\ge Q_2(s):&\text{universal Hawking bound attainable}.
 \end{array}
 \label{eq:three-regions}
\end{equation}
Figure~\ref{fig:phase} shows these three regions in the $(q,\THaw/\omega)$ plane, where $q$ parametrizes the relative composition of the right and left Unruh modes.

Several limits are useful.  As $\THaw/\omega\to\infty$, $s\to1/\sqrt2$ and
\begin{equation}
 q_0=\sqrt{\frac{3+4\sqrt2}{23}}\approx0.6135022146,
 \qquad
 q_2=\sqrt{\frac{3+\sqrt2}{7}}\approx0.7941044878.
 \label{eq:highT-thresholds}
\end{equation}
As $\THaw/\omega\to0^+$,
\begin{equation}
 q_0\to\frac{1}{\sqrt3}\approx0.5773502692,
 \qquad
 q_2\to\sqrt{\frac35}\approx0.7745966692.
 \label{eq:lowT-thresholds}
\end{equation}
The enhancement nevertheless vanishes exponentially in this limit by Eq.~\eqref{eq:lowT-expansion}.

For a fixed $Q=q^2$, Eq.~\eqref{eq:activation-condition} can be inverted.  If $Q\le1/3$, no finite temperature produces strict activation.  If $Q>1/3$, define
\begin{equation}
 s_0(Q)=\frac{1-Q}{5Q-1}.
 \label{eq:s0}
\end{equation}
Whenever $1/\sqrt2<s_0<1$, activation occurs only for
\begin{equation}
 \frac{\THaw}{\omega}
 <\frac{1}{\ln[s_0^2/(1-s_0^2)]}.
 \label{eq:temperature-window}
\end{equation}
For $q>\sqrt{(3+4\sqrt2)/23}$, the inequality is satisfied for every finite positive $\THaw/\omega$ in the fermionic range, although the gain becomes exponentially small at low temperature.

\section{Comparison with state-specific results and physical interpretation}
\label{sec:comparison}

The exact bounds clarify several features that appear only numerically in earlier calculations.  Ref.~\cite{Wu2023} used, among others, the threshold state
\begin{equation}
 \rho_{11}=\sqrt2-1,\quad
 \rho_{22}=\frac12,\quad
 \rho_{33}=\frac{3-2\sqrt2}{2},\quad
 \rho_{44}=\rho_{14}=0,\quad
 \rho_{23}=\frac{\sqrt2-1}{2}.
 \label{eq:ref-state}
\end{equation}
For $q=1$ this state has $f=1/2$ before thermal mixing and its FEF increases with $\THaw$.  At $s=1/\sqrt2$ it reaches
\begin{equation}
 f_{\rm ref}=\frac78-\frac{\sqrt2}{4}
 \approx0.5214466094,
 \label{eq:ref-limit}
\end{equation}
which is below the global value in Eq.~\eqref{eq:highT-fmax}.  Thus the previously observed enhancement is genuine but not optimal.

\begin{figure}[t]
\centering
\includegraphics[width=0.82\textwidth]{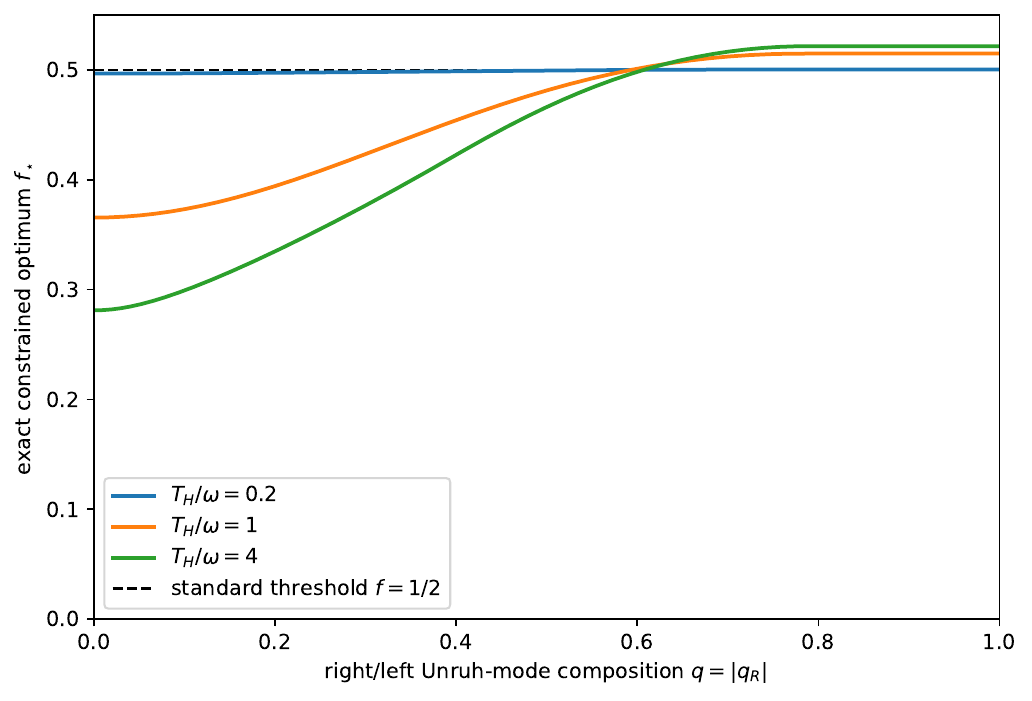}
\caption{Exact constrained optimum $\fstar$ as a function of the relative composition of the right and left Unruh modes for three values of $\THaw/\omega$.  The crossing of $f=1/2$ occurs at the analytic threshold $q_0(s)$; the plateau at larger $q$ is the universal bound.}
\label{fig:fixedq}
\end{figure}

The fixed-composition optimum itself is shown in Fig.~\ref{fig:fixedq}.  For small $q$ the best allowed output stays at or below the standard threshold.  Once $q$ crosses $q_0(s)$, the middle branch becomes strictly activating.  Above $q_2(s)$ the result saturates the universal thermal bound and becomes independent of $q$ because the same accessible optimizer $\sigma_*(s)$ can be reached from different preparations.

The factorization in Eq.~\eqref{eq:factorization} is essential for interpreting these curves.  At $\THaw\to0$, $\HH_\eta$ becomes the identity.  If $q<1$, however, $\LL_q$ remains nontrivial because it encodes which component of the global Unruh excitation is accessible in the detected exterior fermionic mode.  A change caused by varying $q$ at zero temperature is therefore a change of mode preparation or accessibility, not a Hawking-thermal effect.  This point is consistent with the wave-packet analysis of Ref.~\cite{Bruschi2010} and prevents a misleading interpretation of $(q_R,\THaw)$ as two interchangeable noise strengths.

The result also places the Hawking enhancement within the broader quantum-channel literature.  Dissipative activation of teleportation fidelity was known before the black-hole application \cite{Badziag2000,Yeo2008}, and general qubit-channel optimization explains why a nonmaximally entangled input can be optimal \cite{Pal2014}.  What is specific here is the constrained, mode-resolved Schwarzschild problem: the resource is required to remain below the standard FEF threshold immediately before the thermal map, the image restriction at a fixed relative composition of the right and left Unruh modes is retained, and the resulting threshold and optimum are obtained in closed form over the entire two-qubit state space.

Finally, the result is mode resolved.  For a fixed Killing-frequency mode, both the asymptotic frequency and Hawking temperature enter only through $\omega/\THaw$.  A finite-bandwidth detector with a specified local energy gap introduces additional response functions and cannot be replaced by the single-mode map without further assumptions.  The present bounds should therefore be read as exact bounds for the mode model used in the underlying fermionic Schwarzschild construction, not as a detector-level prediction for a specific experiment.

\section{Conclusions}
\label{sec:conclusion}

We have systematically separated Hawking thermal mixing from changes in the relative composition of the right and left Unruh modes in fermionic quantum teleportation near a Schwarzschild horizon.  Starting from the Dirac-field Unruh-mode reduction, we showed that the map seen by the detected exterior fermionic mode factorizes exactly as $\Lambda_{\eta,q}=\HH_\eta\circ\LL_q$.  The amplitudes $q_R$ and $q_L$, constrained by $|q_R|^2+|q_L|^2=1$, specify the relative composition of the prepared right and left Unruh modes, whereas $\eta=[1+\exp(-\omega/\THaw)]^{-1}$ controls the genuine Hawking thermal mixing; these ingredients therefore do not represent interchangeable noise strengths.

On this basis, we solved the constrained global optimization problem at fixed relative composition of the right and left Unruh modes, requiring the accessible two-qubit resource immediately before Hawking mixing to satisfy $f\le1/2$.  A parity twirl, positivity constraints, and monotonicity of the relevant Bell overlap reduce the optimization over arbitrary two-qubit density matrices to a two-variable quadratic problem on the intersection of a disk and a half-plane.  This yields a piecewise analytic optimum, explicit optimizing states, and the necessary-and-sufficient activation condition
\begin{equation}
 q_R^2>\frac{1+\sqrt\eta}{1+5\sqrt\eta}.
\end{equation}
When
\begin{equation}
 q_R^2\ge\frac{1+2\sqrt\eta}{1+4\sqrt\eta},
\end{equation}
the image restriction imposed by the fixed relative composition of the right and left Unruh modes no longer lowers the optimum and the universal Hawking bound is attained.  In the formal high-temperature fermionic limit $\eta=1/2$, this bound becomes
\begin{equation}
 f_{\star}=\frac{9+4\sqrt2}{28}\approx0.5234590803,
 \qquad
 \Ftel=\frac{23+4\sqrt2}{42}\approx0.6823060536.
\end{equation}
In the low-temperature limit, activation may remain kinematically allowed for suitable mode preparations, but its magnitude vanishes exponentially as $\exp(-\omega/\THaw)$.

These results place the previously observed increasing, decreasing, and nonmonotonic state-specific curves within a single bounded quantum-channel optimization problem.  The exact phase diagram contains a strict no-go region, a composition-limited thermal-activation region, and a region in which the universal Hawking bound is attainable.  The main result is therefore an exact channel-level benchmark for Hawking-assisted teleportation: at fixed relative composition of the right and left Unruh modes, the analysis identifies when threshold crossing is impossible, when it is possible but composition limited, and the maximum enhancement allowed by the fermionic Schwarzschild thermal map.

\appendix

\section{Field-mode reduction to the qubit channel}
\label{app:field}

For the massless Dirac mode, introduce $\eta$ as in Eq.~\eqref{eq:eta-s}.  We use the mode ordering (exterior particle, interior antiparticle, exterior antiparticle, interior particle).  In this ordering, the Unruh vacuum can be written as
\begin{equation}
 |0\rangle_U
 =\eta|0000\rangle
 -\sqrt{\eta(1-\eta)}|0101\rangle
 +\sqrt{\eta(1-\eta)}|1010\rangle
 -(1-\eta)|1111\rangle.
 \label{eq:unruh-vacuum-app}
\end{equation}
The one-particle excitation is
\begin{equation}
 \begin{split}
 |1\rangle_U={}&q_R\left[\sqrt\eta|1000\rangle-\sqrt{1-\eta}|1101\rangle\right]\\
 &+q_L\left[\sqrt\eta|0001\rangle+\sqrt{1-\eta}|1011\rangle\right],
 \end{split}
 \label{eq:unruh-one-app}
\end{equation}
with $|q_R|^2+|q_L|^2=1$.  

For a generic qubit density matrix on the Unruh-mode degree of freedom,
\begin{equation}
 \rho_B=
 \begin{pmatrix}x&z\\z^*&y\end{pmatrix},
 \end{equation}
insert Eqs.~\eqref{eq:unruh-vacuum-app} and \eqref{eq:unruh-one-app}, trace the interior modes, and then trace the exterior antifermionic mode when the detector is assumed to respond only to the exterior fermion.  The particle-sector matrix elements are
\begin{equation}
 \rho'_{00}=\eta x+|q_L|^2\eta y,
 \end{equation}
\begin{equation}
 \rho'_{11}=(1-\eta)x+[1-|q_L|^2\eta]y,
 \end{equation}
\begin{equation}
 \rho'_{01}=q_R\sqrt\eta\,z.
\end{equation}
After absorbing the phase of $q_R$ into the detected-mode basis and using $|q_L|^2=1-q^2$, these relations give Eq.~\eqref{eq:full-map}.

\section{Reduction of the global optimization}
\label{app:reduction}

This appendix proves the reduction underlying Eqs.~\eqref{eq:universal-bound} and \eqref{eq:xy-domain}.

Let the maximally entangled state that attains the output FEF be $|\Phi_U\rangle=(U_A\otimes\id)|\Phi^+\rangle$.  A unitary on Alice commutes with both Bob-side channels, so we may absorb $U_A$ into the preparation and optimize the $|\Phi^+\rangle$ overlap.  Now apply the parity twirl
\begin{equation}
 \mathcal{T}(\rho)=\frac12\left[\rho+(Z\otimes Z)\rho(Z\otimes Z)\right].
 \label{eq:parity-twirl}
\end{equation}
The maps $\LL_q$ and $\HH_\eta$ are $Z$ covariant, and $(Z\otimes Z)|\Phi^+\rangle=|\Phi^+\rangle$.  Therefore the target output overlap is unchanged.  Since FEF is convex and invariant under local unitaries,
\begin{equation}
 f[\mathcal{T}(\sigma)]\le f(\sigma),
\end{equation}
so the threshold constraint is preserved.  Hence an optimizer exists among $X$ states.

Write the preparation state as
\begin{equation}
 \rho=
 \begin{pmatrix}
 A&0&0&W\\
 0&B&Z&0\\
 0&Z^*&C&0\\
 W^*&0&0&D
 \end{pmatrix}.
 \label{eq:generic-X-app}
\end{equation}
For the $|\Phi^+\rangle$ target, $Z$ does not enter the objective.  Setting $Z=0$ cannot increase any competing Bell-sector overlap, so it preserves the FEF constraint while leaving the target objective unchanged.  Next replace
\begin{equation}
 A\mapsto A+(1-q^2)B,
 \qquad
 C\mapsto C+q^2B,
 \qquad
 B\mapsto0,
 \label{eq:B-eliminate}
\end{equation}
while leaving $D$ and $W$ unchanged.  This transformation preserves the trace.  Positivity is also preserved: the $A$ entry of the $|00\rangle$--$|11\rangle$ block only increases, while the new $C$ remains nonnegative.  It also preserves both Bell-sector constraints on the FEF after $\LL_q$ and increases the final $|\Phi^+\rangle$ overlap by $(1-s^2)q^2B/2\ge0$.  Thus $B=0$ can be imposed without loss.

With $B=Z=0$, positivity gives $|W|^2\le AD$.  The output objective increases with $|W|$.  If the FEF constraint is inactive, positivity is therefore saturated.  If the FEF constraint is active first, then
\begin{equation}
 A+q^2D+2q|W|=1.
 \label{eq:f-active}
\end{equation}
Substitution into the output objective shows that its coefficients of $A$ and $D$ are $(2s+1)(s-1)$ and $s(s-1)q^2$, respectively, both nonpositive for $s\le1$.  Any point with $|W|^2<AD$ can therefore be moved toward smaller $A$ or $D$ while increasing $|W|$ until positivity saturates, without reducing the objective.  Consequently an optimizer exists with
\begin{equation}
 |W|=\sqrt{AD}.
 \label{eq:W-saturate}
\end{equation}
Choose the phase so that $W\ge0$ and write $A=x^2$, $D=y^2$.  Normalization gives $x^2+y^2\le1$, while Eq.~\eqref{eq:f-active} becomes $(x+qy)^2\le1$.  The output overlap is then Eq.~\eqref{eq:xy-objective}.  This proves the exact two-variable reduction.

For the universal thermal problem there is no image restriction.  Setting $q=1$ in the line part of this reduction gives Eq.~\eqref{eq:univ-parabola} and hence Theorem~\ref{thm:universal}.

\section{Branch boundaries and their ordering}
\label{app:branches}

On the circular boundary $x^2+y^2=1$, Eq.~\eqref{eq:xy-objective} is determined by the symmetric matrix
\begin{equation}
 M=
 \begin{pmatrix}
 2s^2-1&s\sqrt Q\\
 s\sqrt Q&s^2Q
 \end{pmatrix}.
 \label{eq:M-app}
\end{equation}
Its largest eigenvalue gives Eq.~\eqref{eq:fI}.  The intersection of the circle and line is Eq.~\eqref{eq:circle-line-point}.  Requiring this point to be stationary on the circle gives
\begin{equation}
 (2s^2-s)Q^2+(-6s^2+6s+2)Q+(4s^2-s-2)=0.
 \label{eq:Q1-poly}
\end{equation}
The physical root is Eq.~\eqref{eq:Q1}.  It becomes zero when $4s^2-s-2=0$, which yields $s_c=(1+\sqrt{33})/8$.

On the straight boundary $x+qy=1$, substitution $y=(1-x)/q$ gives
\begin{equation}
 2f_{\rm out}(x)
 =1+3s^2x^2-2s^2x-2sx^2+2sx-x^2.
 \label{eq:line-parabola-app}
\end{equation}
Its stationary point is Eq.~\eqref{eq:line-stationary}.  The condition $x_*^2+y_*^2\le1$ reduces to
\begin{equation}
 Q\ge\frac{1+2s}{1+4s},
\end{equation}
which proves Eq.~\eqref{eq:Q2}.

The activation boundary follows directly from Eq.~\eqref{eq:fII-factor}.  To prove the ordering, first
\begin{equation}
 Q_2-Q_0
 =\frac{2s(3s+1)}{(4s+1)(5s+1)}>0.
 \label{eq:Q2-Q0}
\end{equation}
When $Q_1$ exists, denote the left-hand side of Eq.~\eqref{eq:Q1-poly} by $P(Q)$.  Direct substitution gives
\begin{equation}
 P(Q_0)=
 \frac{4s(2s-1)(3s+1)^2}{(5s+1)^2}>0.
 \label{eq:PQ0}
\end{equation}
In the same range $P(0)=4s^2-s-2<0$, so the positive physical root satisfies $Q_1<Q_0$.  This proves Eq.~\eqref{eq:ordering}.

\section*{Funding}
No funding was received for this work.

\section*{Conflict of interest}
The author declares no conflict of interest.

\section*{Data availability}
No datasets were generated or analyzed in this study.

\bibliographystyle{apsrev4-2}
\bibliography{refs}

\begin{thebibliography}{15}%
\makeatletter
\providecommand \@ifxundefined [1]{%
 \@ifx{#1\undefined}
}%
\providecommand \@ifnum [1]{%
 \ifnum #1\expandafter \@firstoftwo
 \else \expandafter \@secondoftwo
 \fi
}%
\providecommand \@ifx [1]{%
 \ifx #1\expandafter \@firstoftwo
 \else \expandafter \@secondoftwo
 \fi
}%
\providecommand \natexlab [1]{#1}%
\providecommand \enquote  [1]{``#1''}%
\providecommand \bibnamefont  [1]{#1}%
\providecommand \bibfnamefont [1]{#1}%
\providecommand \citenamefont [1]{#1}%
\providecommand \href@noop [0]{\@secondoftwo}%
\providecommand \href [0]{\begingroup \@sanitize@url \@href}%
\providecommand \@href[1]{\@@startlink{#1}\@@href}%
\providecommand \@@href[1]{\endgroup#1\@@endlink}%
\providecommand \@sanitize@url [0]{\catcode `\\12\catcode `\$12\catcode `\&12\catcode `\#12\catcode `\^12\catcode `\_12\catcode `\%12\relax}%
\providecommand \@@startlink[1]{}%
\providecommand \@@endlink[0]{}%
\providecommand \url  [0]{\begingroup\@sanitize@url \@url }%
\providecommand \@url [1]{\endgroup\@href {#1}{\urlprefix }}%
\providecommand \urlprefix  [0]{URL }%
\providecommand \Eprint [0]{\href }%
\providecommand \doibase [0]{https://doi.org/}%
\providecommand \selectlanguage [0]{\@gobble}%
\providecommand \bibinfo  [0]{\@secondoftwo}%
\providecommand \bibfield  [0]{\@secondoftwo}%
\providecommand \translation [1]{[#1]}%
\providecommand \BibitemOpen [0]{}%
\providecommand \bibitemStop [0]{}%
\providecommand \bibitemNoStop [0]{.\EOS\space}%
\providecommand \EOS [0]{\spacefactor3000\relax}%
\providecommand \BibitemShut  [1]{\csname bibitem#1\endcsname}%
\let\auto@bib@innerbib\@empty
\bibitem [{\citenamefont {Bennett}\ \emph {et~al.}(1993)\citenamefont {Bennett}, \citenamefont {Brassard}, \citenamefont {Cr{\'e}peau}, \citenamefont {Jozsa}, \citenamefont {Peres},\ and\ \citenamefont {Wootters}}]{Bennett1993}%
  \BibitemOpen
  \bibfield  {author} {\bibinfo {author} {\bibfnamefont {C.~H.}\ \bibnamefont {Bennett}}, \bibinfo {author} {\bibfnamefont {G.}~\bibnamefont {Brassard}}, \bibinfo {author} {\bibfnamefont {C.}~\bibnamefont {Cr{\'e}peau}}, \bibinfo {author} {\bibfnamefont {R.}~\bibnamefont {Jozsa}}, \bibinfo {author} {\bibfnamefont {A.}~\bibnamefont {Peres}},\ and\ \bibinfo {author} {\bibfnamefont {W.~K.}\ \bibnamefont {Wootters}},\ }\href@noop {} {\bibfield  {journal} {\bibinfo  {journal} {Phys. Rev. Lett.}\ }\textbf {\bibinfo {volume} {70}},\ \bibinfo {pages} {1895} (\bibinfo {year} {1993})}\BibitemShut {NoStop}%
\bibitem [{\citenamefont {Horodecki}\ \emph {et~al.}(1999)\citenamefont {Horodecki}, \citenamefont {Horodecki},\ and\ \citenamefont {Horodecki}}]{Horodecki1999}%
  \BibitemOpen
  \bibfield  {author} {\bibinfo {author} {\bibfnamefont {M.}~\bibnamefont {Horodecki}}, \bibinfo {author} {\bibfnamefont {P.}~\bibnamefont {Horodecki}},\ and\ \bibinfo {author} {\bibfnamefont {R.}~\bibnamefont {Horodecki}},\ }\href@noop {} {\bibfield  {journal} {\bibinfo  {journal} {Phys. Rev. A}\ }\textbf {\bibinfo {volume} {60}},\ \bibinfo {pages} {1888} (\bibinfo {year} {1999})}\BibitemShut {NoStop}%
\bibitem [{\citenamefont {Badzi{\k{a}}g}\ \emph {et~al.}(2000)\citenamefont {Badzi{\k{a}}g}, \citenamefont {Horodecki}, \citenamefont {Horodecki},\ and\ \citenamefont {Horodecki}}]{Badziag2000}%
  \BibitemOpen
  \bibfield  {author} {\bibinfo {author} {\bibfnamefont {P.}~\bibnamefont {Badzi{\k{a}}g}}, \bibinfo {author} {\bibfnamefont {M.}~\bibnamefont {Horodecki}}, \bibinfo {author} {\bibfnamefont {P.}~\bibnamefont {Horodecki}},\ and\ \bibinfo {author} {\bibfnamefont {R.}~\bibnamefont {Horodecki}},\ }\href@noop {} {\bibfield  {journal} {\bibinfo  {journal} {Phys. Rev. A}\ }\textbf {\bibinfo {volume} {62}},\ \bibinfo {pages} {012311} (\bibinfo {year} {2000})}\BibitemShut {NoStop}%
\bibitem [{\citenamefont {Yeo}(2008)}]{Yeo2008}%
  \BibitemOpen
  \bibfield  {author} {\bibinfo {author} {\bibfnamefont {Y.}~\bibnamefont {Yeo}},\ }\href@noop {} {\bibfield  {journal} {\bibinfo  {journal} {Phys. Rev. A}\ }\textbf {\bibinfo {volume} {78}},\ \bibinfo {pages} {022334} (\bibinfo {year} {2008})}\BibitemShut {NoStop}%
\bibitem [{\citenamefont {Fuentes-Schuller}\ and\ \citenamefont {Mann}(2005)}]{FuentesSchuller2005}%
  \BibitemOpen
  \bibfield  {author} {\bibinfo {author} {\bibfnamefont {I.}~\bibnamefont {Fuentes-Schuller}}\ and\ \bibinfo {author} {\bibfnamefont {R.~B.}\ \bibnamefont {Mann}},\ }\href@noop {} {\bibfield  {journal} {\bibinfo  {journal} {Phys. Rev. Lett.}\ }\textbf {\bibinfo {volume} {95}},\ \bibinfo {pages} {120404} (\bibinfo {year} {2005})}\BibitemShut {NoStop}%
\bibitem [{\citenamefont {Pan}\ and\ \citenamefont {Jing}(2008)}]{PanJing2008}%
  \BibitemOpen
  \bibfield  {author} {\bibinfo {author} {\bibfnamefont {Q.}~\bibnamefont {Pan}}\ and\ \bibinfo {author} {\bibfnamefont {J.}~\bibnamefont {Jing}},\ }\href@noop {} {\bibfield  {journal} {\bibinfo  {journal} {Phys. Rev. D}\ }\textbf {\bibinfo {volume} {78}},\ \bibinfo {pages} {065015} (\bibinfo {year} {2008})}\BibitemShut {NoStop}%
\bibitem [{\citenamefont {Mart{\'i}n-Mart{\'i}nez}\ \emph {et~al.}(2010)\citenamefont {Mart{\'i}n-Mart{\'i}nez}, \citenamefont {Garay},\ and\ \citenamefont {Le{\'o}n}}]{MartinMartinez2010}%
  \BibitemOpen
  \bibfield  {author} {\bibinfo {author} {\bibfnamefont {E.}~\bibnamefont {Mart{\'i}n-Mart{\'i}nez}}, \bibinfo {author} {\bibfnamefont {L.~J.}\ \bibnamefont {Garay}},\ and\ \bibinfo {author} {\bibfnamefont {J.}~\bibnamefont {Le{\'o}n}},\ }\href@noop {} {\bibfield  {journal} {\bibinfo  {journal} {Phys. Rev. D}\ }\textbf {\bibinfo {volume} {82}},\ \bibinfo {pages} {064006} (\bibinfo {year} {2010})}\BibitemShut {NoStop}%
\bibitem [{\citenamefont {Bruschi}\ \emph {et~al.}(2010)\citenamefont {Bruschi}, \citenamefont {Louko}, \citenamefont {Mart{\'i}n-Mart{\'i}nez}, \citenamefont {Dragan},\ and\ \citenamefont {Fuentes}}]{Bruschi2010}%
  \BibitemOpen
  \bibfield  {author} {\bibinfo {author} {\bibfnamefont {D.~E.}\ \bibnamefont {Bruschi}}, \bibinfo {author} {\bibfnamefont {J.}~\bibnamefont {Louko}}, \bibinfo {author} {\bibfnamefont {E.}~\bibnamefont {Mart{\'i}n-Mart{\'i}nez}}, \bibinfo {author} {\bibfnamefont {A.}~\bibnamefont {Dragan}},\ and\ \bibinfo {author} {\bibfnamefont {I.}~\bibnamefont {Fuentes}},\ }\href@noop {} {\bibfield  {journal} {\bibinfo  {journal} {Phys. Rev. A}\ }\textbf {\bibinfo {volume} {82}},\ \bibinfo {pages} {042332} (\bibinfo {year} {2010})}\BibitemShut {NoStop}%
\bibitem [{\citenamefont {Pal}\ \emph {et~al.}(2014)\citenamefont {Pal}, \citenamefont {Bandyopadhyay},\ and\ \citenamefont {Ghosh}}]{Pal2014}%
  \BibitemOpen
  \bibfield  {author} {\bibinfo {author} {\bibfnamefont {R.}~\bibnamefont {Pal}}, \bibinfo {author} {\bibfnamefont {S.}~\bibnamefont {Bandyopadhyay}},\ and\ \bibinfo {author} {\bibfnamefont {S.}~\bibnamefont {Ghosh}},\ }\href@noop {} {\bibfield  {journal} {\bibinfo  {journal} {Phys. Rev. A}\ }\textbf {\bibinfo {volume} {90}},\ \bibinfo {pages} {052304} (\bibinfo {year} {2014})}\BibitemShut {NoStop}%
\bibitem [{\citenamefont {Banerjee}\ \emph {et~al.}(2017)\citenamefont {Banerjee}, \citenamefont {Alok}, \citenamefont {Omkar},\ and\ \citenamefont {Srikanth}}]{Banerjee2017}%
  \BibitemOpen
  \bibfield  {author} {\bibinfo {author} {\bibfnamefont {S.}~\bibnamefont {Banerjee}}, \bibinfo {author} {\bibfnamefont {A.~K.}\ \bibnamefont {Alok}}, \bibinfo {author} {\bibfnamefont {S.}~\bibnamefont {Omkar}},\ and\ \bibinfo {author} {\bibfnamefont {R.}~\bibnamefont {Srikanth}},\ }\href@noop {} {\bibfield  {journal} {\bibinfo  {journal} {JHEP}\ }\textbf {\bibinfo {volume} {02}},\ \bibinfo {pages} {082}}\BibitemShut {NoStop}%
\bibitem [{\citenamefont {Chen}\ and\ \citenamefont {Chan}(2019)}]{ChenChan2019}%
  \BibitemOpen
  \bibfield  {author} {\bibinfo {author} {\bibfnamefont {X.}~\bibnamefont {Chen}}\ and\ \bibinfo {author} {\bibfnamefont {K.~W.~C.}\ \bibnamefont {Chan}},\ }\href@noop {} {\bibfield  {journal} {\bibinfo  {journal} {Phys. Rev. A}\ }\textbf {\bibinfo {volume} {99}},\ \bibinfo {pages} {022334} (\bibinfo {year} {2019})}\BibitemShut {NoStop}%
\bibitem [{\citenamefont {Wu}\ \emph {et~al.}(2023)\citenamefont {Wu}, \citenamefont {Fan}, \citenamefont {Wang}, \citenamefont {Wu}, \citenamefont {Huang},\ and\ \citenamefont {Zeng}}]{Wu2023}%
  \BibitemOpen
  \bibfield  {author} {\bibinfo {author} {\bibfnamefont {S.-M.}\ \bibnamefont {Wu}}, \bibinfo {author} {\bibfnamefont {X.-W.}\ \bibnamefont {Fan}}, \bibinfo {author} {\bibfnamefont {R.-D.}\ \bibnamefont {Wang}}, \bibinfo {author} {\bibfnamefont {H.-Y.}\ \bibnamefont {Wu}}, \bibinfo {author} {\bibfnamefont {X.-L.}\ \bibnamefont {Huang}},\ and\ \bibinfo {author} {\bibfnamefont {H.-S.}\ \bibnamefont {Zeng}},\ }\href@noop {} {\bibfield  {journal} {\bibinfo  {journal} {JHEP}\ }\textbf {\bibinfo {volume} {11}},\ \bibinfo {pages} {232}}\BibitemShut {NoStop}%
\bibitem [{\citenamefont {Mi}\ \emph {et~al.}(2025)\citenamefont {Mi}, \citenamefont {Huang}, \citenamefont {Fei},\ and\ \citenamefont {Zhang}}]{Mi2025}%
  \BibitemOpen
  \bibfield  {author} {\bibinfo {author} {\bibfnamefont {G.-W.}\ \bibnamefont {Mi}}, \bibinfo {author} {\bibfnamefont {X.}~\bibnamefont {Huang}}, \bibinfo {author} {\bibfnamefont {S.-M.}\ \bibnamefont {Fei}},\ and\ \bibinfo {author} {\bibfnamefont {T.}~\bibnamefont {Zhang}},\ }\href@noop {} {\bibfield  {journal} {\bibinfo  {journal} {Annalen Phys.}\ }\textbf {\bibinfo {volume} {537}},\ \bibinfo {pages} {2400308} (\bibinfo {year} {2025})}\BibitemShut {NoStop}%
\bibitem [{\citenamefont {Liu}\ \emph {et~al.}(2025)\citenamefont {Liu}, \citenamefont {Long},\ and\ \citenamefont {He}}]{LiuLongHe2025}%
  \BibitemOpen
  \bibfield  {author} {\bibinfo {author} {\bibfnamefont {C.-y.}\ \bibnamefont {Liu}}, \bibinfo {author} {\bibfnamefont {Z.-w.}\ \bibnamefont {Long}},\ and\ \bibinfo {author} {\bibfnamefont {Q.-l.}\ \bibnamefont {He}},\ }\href@noop {} {\bibfield  {journal} {\bibinfo  {journal} {Eur. Phys. J. C}\ }\textbf {\bibinfo {volume} {85}},\ \bibinfo {pages} {926} (\bibinfo {year} {2025})}\BibitemShut {NoStop}%
\bibitem [{\citenamefont {Liu}\ and\ \citenamefont {Li}(2025)}]{LiuLi2025}%
  \BibitemOpen
  \bibfield  {author} {\bibinfo {author} {\bibfnamefont {L.}~\bibnamefont {Liu}}\ and\ \bibinfo {author} {\bibfnamefont {L.}~\bibnamefont {Li}},\ }\href@noop {} {\bibfield  {journal} {\bibinfo  {journal} {Quantum Inf. Process.}\ }\textbf {\bibinfo {volume} {24}},\ \bibinfo {pages} {1} (\bibinfo {year} {2025})}\BibitemShut {NoStop}%
\end{thebibliography}%

\end{document}